# SIP OVERLOAD CONTROL TESTBED: DESIGN, BUILDING AND EVALUATION

Ahmad reza Montazerolghaem[1], Mohammad Hossein Yaghmaee[2]

[1,2]Department of Computer Engineering, Ferdowsi University of mashhad, Mashhad, Iran

`Ahmadreza.montazerolghaem@stu-mail.um.ac.ir`
`yaghmaee@ieee.org`

## ABSTRACT

*Having facilities such as being in text form, end-to-end connection establishment, and being independence from the type of transmitted data, SIP protocol is a good choice for signaling protocol in order to set up a connection between two users of an IP network. Although utilization of SIP protocol in a wide range of applications has made various vulnerabilities in this protocol, amongst which overload could make serious problems in SIP servers. A SIP is overloaded when it does not have sufficient resources (majorly CPU processing power and memory) to process all messages. In this paper the window-based overload control mechanism which does not require explicit feedback is developed and implemented on Asterisk open source proxy and evaluated. The results of implementation show that this method could practically maintain throughput in case of overload. As we know this is the only overload control method which is implemented on a real platform without using explicit feedback. The results show that the under load server maintains its throughput at the maximum capacity.*

## 1. INTRODUCTION

SIP server is an application one. The overload problem in SIP server is distinguished with ones in other HTTP servers for at least three reasons: firstly the messages of SIP meeting pass several SIP proxy servers to reach destination which itself could make overload between two SIP proxy servers. Secondly SIP has several retransmit timers which are used for dealing with package loss, especially when the package is sent via UDP transmission protocol, and this could lead to overload on SIP proxy server. Thirdly SIP requests are used as real time session signaling, so have a high sensitivity. Overload in SIP-based networks occur when the server does not have sources necessary for answering every received call. Reviews accomplished in overloaded SIP proxy server show that increasing request rate results in sudden increase in delay in establishing connection and dropping proxy throughput and therefore increase in unsuccessful call rates. Therefore the aim in overload control in SIP is maintaining the throughput of overloaded server near its capacity. Generally there are two local and distributed methods for overload control. In local control when SIP proxy server reaches its capacity threshold, it starts to reject requests; SIP estimates this threshold by calculating CPU consumption or queue length. But request rejection mechanism in order to finish meeting imposes cost itself and when server is overloaded it is compelled to allocate fraction of sources to reject requests, which in turn decreases efficiency in SIP proxy server. In distributed method, upstream servers control the load of downstream servers through rejecting requests and try to maintain it under their capacity. Figure 1 illustrates connection establishment between two user agents in a case in

17

International Journal of Ambient Systems and Applications (IJASA) Vol.1, No.2, June 2013

which middle proxies are configured statefully. The proxy task is routing and redeploying signaling between user agents.

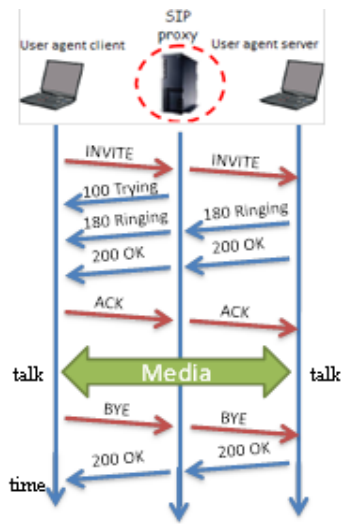

Figure 1. Exchanged messages for establishing connection in SIP

## 2. RELATED WORK

Many researches about the efficiency of SIP proxy server have been accomplished. Paper [1] deals with overload control methods in SIP proxy server and uses OPNET software for measuring throughput. In papers [2] and [3] SIP is practically implemented along with TCP and UDP transmission protocol and OpenSER is used to obtain efficiency results. Articles [4] and [5] mention to window-based distributed method and combination of signal and window-based method, respectively. SIPstone [6] is series of benchmark in which various criterions are proposed for evaluating proxy server powers, redirect server and registrar in answering SIP requests. In [7] another benchmark is presented for measuring the effect of operating system, hardware configuration, database, and selected transmission layer on SIP efficiency. In [8] practical experiments are accomplished on four types of proxy implementation which are different in both thread management and memory allocation method. The results of these experiments show that the effective parameters in proxy efficiency could be classified in two parts: parameters related to protocol such as message length, length variability, and irregularity of excess load, and parameters related to the type of server implementation e.g. how to allocate sources of operating system to transactions. Also in [9] similar studies about the effect of operating system and type of proxy implementation on SIP efficiency are done. In [10] it is studied about the efficiency of signaling of SIP in establishing VOIP connections using JAIN SIP API and by considering the effect of call duration and call rate on the delay of connection establishment between two end-to-end user agents. In [11] the effect of delay of user's answer on SIP server's efficiency is analyzed by introducing a tool called SIPperformer. In [12] issues such importance of security cost in configuration which uses authentication and the effect of proxy being stateful or stateless and the protocol type of transmission layer on proxy's efficiency is studied by placing only one proxy between user agents. A cluster of researches is concentrated on the evaluation of SIP efficiency under various access technologies. For example in [13] the effect of transmission delay and package loss on W-CDMA link is surveyed by using a proxy. In [14] the delay in signaling of SIP in establishing IMS meetings is evaluated for different WiMax channels with different speeds. Also compression techniques for SIP messages are used in order to reduce the volume of SIP packages along with their transmission delay.

Selection of transmission layer protocol is also influential in the efficiency of signaling of SIP. In [15] various options in selection of transmission layer protocol for SIP are surveyed qualitatively. In [16] the effect of deploying various transmission layer protocols, especially effect of window control mechanism in TCP on throughput and delay in connection establishment is evaluated. In [17] it is shown that despite the general perception in which the more common utilization of UDP than TCP is considered on account of the low processing excess load in the former, it is probable that unfavorable efficiency in TCP utilization is due to implementation manner of proxy.

In this paper the overload control method is implemented on an appropriate platform and since this method does not need any feedback, it is very robust. This method is one of window-based techniques which is developed and implemented on Asterisk open source proxy and evaluated. In the following in parts II and III this open source proxy and the implementation platform are introduced, respectively. In part IV the details of load control algorithm which is developed in this open source software is presented. Part V includes the results and evaluation of efficiency of implemented algorithm and finally part VI includes conclusion and topics of future works.

## 3. INTRODUCING ASTERISK PROXY

Asterisk is the most popular open source VOIP telephone system in the world and currently many available IPPBXs are produced on its basis. Asterisk is based on C programming language and could be loaded in various operating systems such as Linux NetBSD, UNIX, Solaris, and Mac OSX. In addition it is observed that some versions of Asterisk are installable and operable in Windows platform. Although Asterisk services could be operable by using common computers and servers and calculating power of system (CPU/RAM) on the basis of users multiplicity, the popularity of Asterisk and diversity of its services have made producers to utilize many of combined platforms of Linux and Asterisk in producing integrated connection equipments in various scales. The minimum system requirements for installing Asterisk are a 500 MHz Pentium computer with 512 MB RAM and 20 GB empty hard space [37]. This software uses UDP and TCP transmission protocols to receive and send SIP messages, and when receiving SIP messages it first intercepts the message and then decides whether to reply it or forward it to next destination [35, 36]. In this paper UDP transmission protocol is used to receive and send SIP messages.

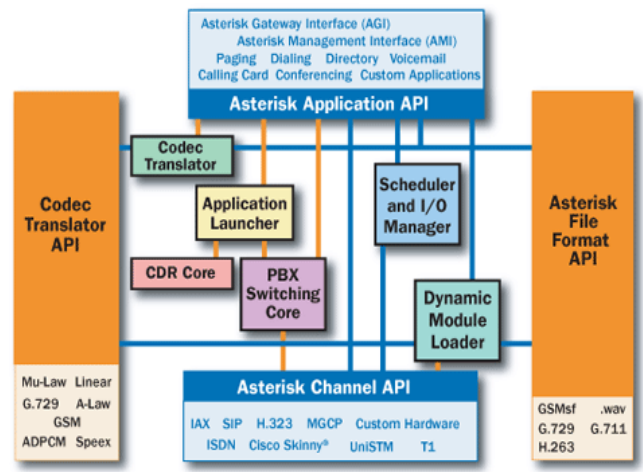

Figure 2. Asterisk Architecture

Asterisk uses several Worker Processes to receive and send SIP messages and every Worker Process receives message individually and makes decision about it. In order to process a SIP message, Worker Process should make a connection between the message and the transaction; the message could relate to a transaction which already existed or it may be new message for which a transaction is created; these transactions are saved in shared memory of Worker Processes. There is no guarantee that a Worker Process manages every messages related to same transaction and it is probable that one transaction be managed by several Worker Processes. When a message is sent, a new timer is created and added to a list. A process manages this list, checks timers, and until timer finishes and no replication for that message has been received, resends the message by accessing the appropriate transaction [34].

## 4. INTRODUCING IMPLEMENTATION PLATFORM

The topology which is used to evaluate and apply overload control methods is a SIP trapezoidal structure which is shown in figure 5. In this topology, the messages of each sides of call are exchanged via two middle proxies. In this topology it is assumed that M transmitter or upstream proxies (e.g. M=1) make an overload in a destination (downstream) proxy by sending many call-making requests. The overload which the proxy in this topology faces with is in the form of server-server. In this form of overload a limited number of upstream proxies send a huge volume of traffic to a downstream proxy and leads it to face with overload. This topology is applicable wherever any users get service from their local service provider proxy. The capacities of upstream proxies are considered so that they do not face with overload during experiments. By the way for simplicity only one upstream proxy is used. Asterisk software and Spirent Abacus 5000 tester device are used for implementing proxy servers and user agents, respectively. The upstream server is a PC with INTEL Dual Core 2.8 GHZ processor and 4 GB memory and the downstream server is a PC with INTEL 1.8 GHz processor and 2 GB memory. Both servers uses version 6.3 of Linux CentOS as their operating system. We used Spirent Abacus 5000 device to create traffic with different transmission rates and various distributions. This device is used for different tests including interoperability, performance, scalability, and testing audio and video quality on IP networks. This production is able to test the efficiency and extensibility of tested proxy by producing hundreds to thousands of calls.

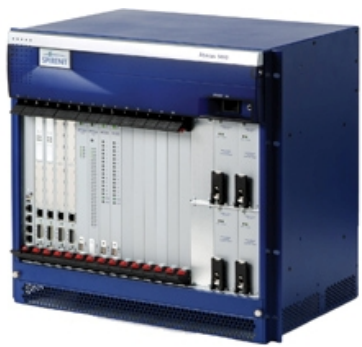

Figure 3. Spirent Abacus 5000 device

International Journal of Ambient Systems and Applications (IJASA) Vol.1, No.2, June 2013

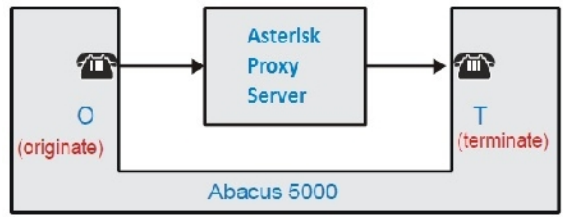

Figure 4. the role of Spirent Abacus 5000 device

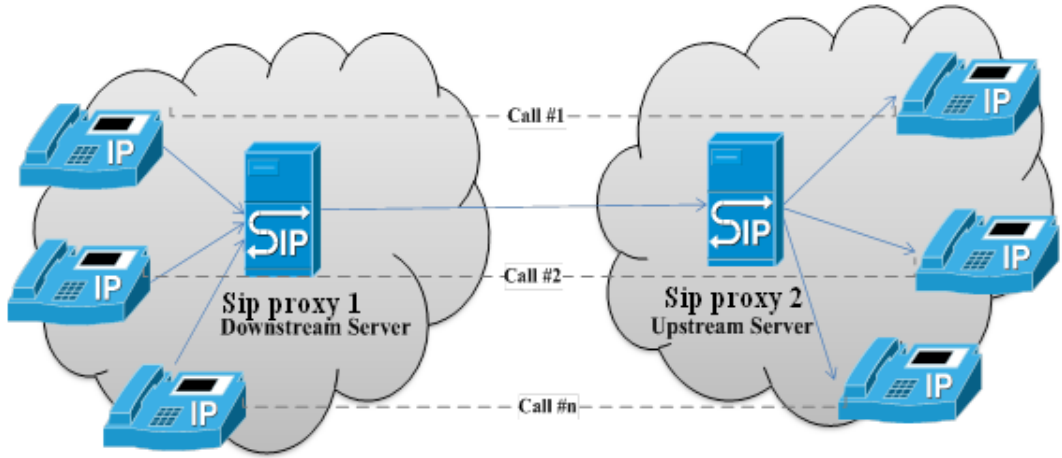

Figure 5. Dual-Proxy Topology

## 5. OVERLOAD CONTROL ALGORITHM

In this algorithm the upstream server keeps a window of active transactions. New calls will be accepted if only the window has any empty space. When a transaction finishes, its delay is calculated and added to the list of recent transactions' delays; then the average of recent delays is calculated and if the amount of this average is more than multiplication of alpha by momentary delay, the magnitude of window will be equal to one and the window potential size, which we call WINth afterwards, reduces to half of the window size. The threshold of WINth accelerates the proceeding of window increase so that if the window size is lower than WINth, window size will be increase by one unit; and if it is higher than that, it will increase by amount of 1/W [33]. The pseudo-code of how window size will change during establishing a connection is as follows.

If $Z_{avg}$ $Z_{th}$ + × then Windows$t+1$=1, Else if Windows$t$ < WINth then Windows$t+1$=Windows$t$ + 1, Else if Windows$t$ WINth then Windows$t+1$=Windows$t$ + 1/ Windows$t$

$Z_{avg}$ is the average of delays in recent transactions and parameter $Z_{th}$ describes the acceptable delay threshold. is the variance of delays and is a parameter for regulating acceptable variations of average delay about the given threshold $Z_{th}$ and is considered equal to 3 practically [32].

## 6. EVALUATION OF EFFICIENCY

When calls are outbound, the rout of signaling is between the proxies related to each domain. signaling load is produced by two UAS and UAC user agents, both of which role is played by

21

Spirent Abacus 5000 device. Conversation production rate starts from low amount and continues to heavy rates of about 1800 cps. According to figure 6, whenever an outbound request reaches proxy, the proxy firstly finds the IP address of proxy of destination network through sending DNS request and then sends the request to it. Calling with DNS will increase the delay in call establishment time a little. One solution is to keep the addresses of active neighbor proxies in the proxy's cache. Of course this method is not appropriate when DNS is used for distributing load between different proxies [20].

On account of the results of last section, the capacity of downstream proxy under load in platform of experiment in figure 4 is set to 700 cps, link delay and probability of package loss are negligible generally, and magnitude of is considered 1.2 [33].

Figure 7 shows the throughput as a function of rate of received call requests in case of existence and non-existence of overload control method, which represents that proxy's throughput could maintain at its maximum capacity in case of existence of overload control mechanism.

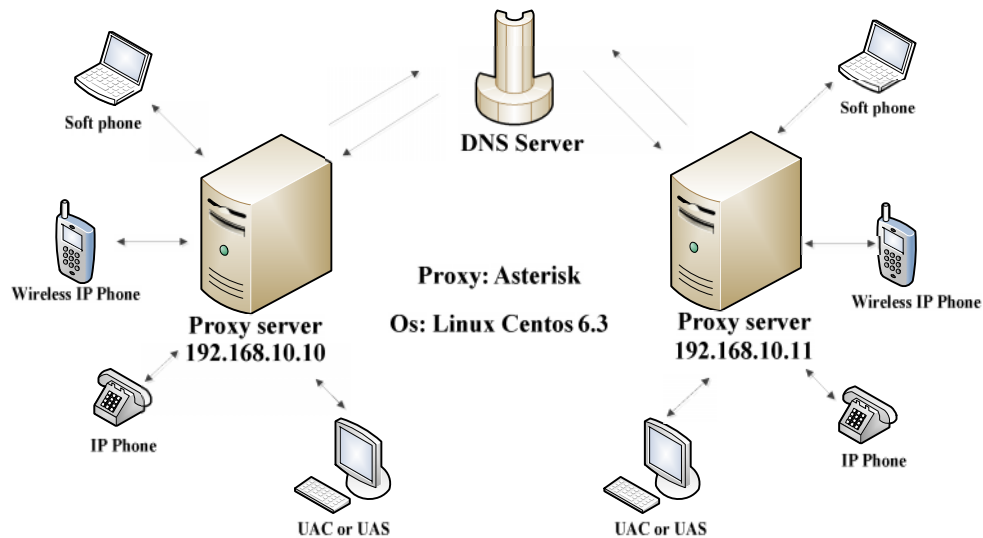

Figure 6. Dual-Proxy Topology

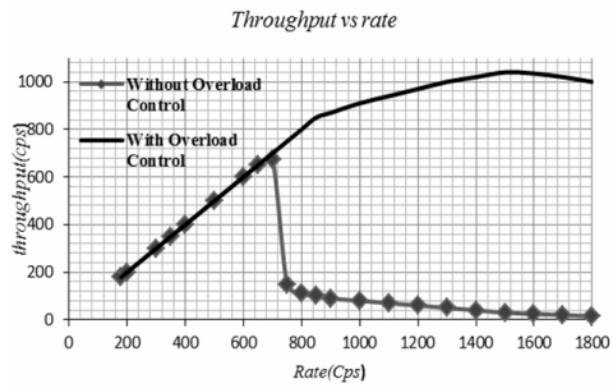

Figure 7. Throughput in case of existence and non-existence of overload control mechanism

The size of window increases along with receiving new requests, so does delay. When the average of resulted delays is lower than multiplication of in momentary delay, size of window reduces to one and WINth reduces to half of the size of window and then starts to increase again. As it is shown in figures 7 and 8, this makes upstream proxy, in which overload control mechanism is implemented, maintain its throughput on about twice of downstream proxy's capacity, and also increases the average time of call establishment in this proxy to about 1500 cps linearly and with a growth rate far much lower than the case in which the overload control mechanism is not used. In rates higher than downstream server's capacity (700 cps), the huge amount of received requests stimulate CPU sensor and therefore many calls are rejected. Sudden rejection of calls leads to many Ack packages reach proxy in a very short interval and therefore full the queue of proxy so that there is no place in this queue for the packages of answers related to ongoing calls. Missing of answer packages is a stimulation to activate resend mechanisms in both server and client which deteriorates the situation. So calls accepted in proxy in this status are accepted with a very long delay.

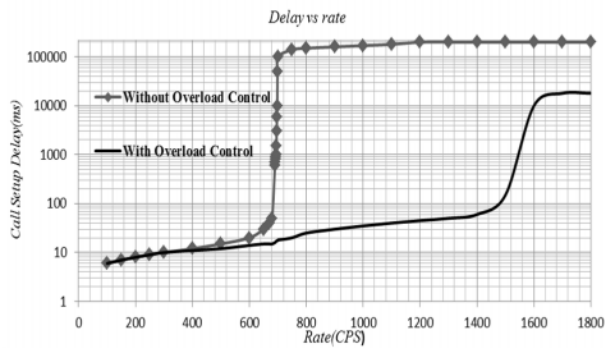

Figure 8. Average delay of call establishment in presence and Absence of overload control mechanism

The diagram shown in figure 9 illustrates resend rate for INVITE and BYE requests from user side, individually. As it is expected, when we use overload control mechanism in upstream server, resend rates of messages decrease considerably. Overload leads to loss of OK packages related to the passed calls. So the proxy is required to resend INVITE requests related to lost packages. In this case, increase of resend rate makes proxy spend much of its time on resending requests related to ongoing calls and therefore throughput rate of proxy falls considerably. Processing the abundant packages which exist in proxy's queue, delays the passes calls more and increases the resend rate in caller's side.

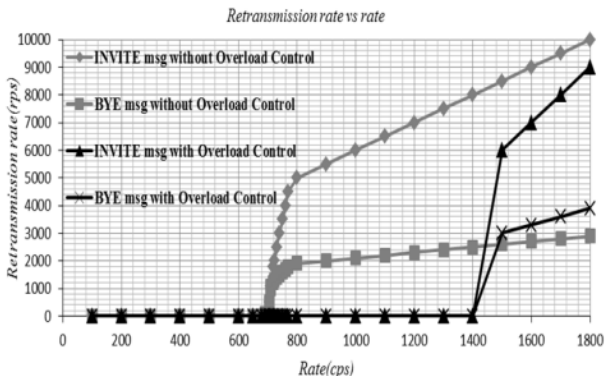

Figure 9. Resend rate of INVITE and BYE messages in presence and absence of overload control mechanism

The information concluded from above diagrams is summarized in below charts. As it is clear in charts 1 and 2, improvement of efficiency in Asterisk proxy utilizing overload control mechanism is in case of congestion, so that improves throughput parameters and CPU and memory utilization.

Table 1: Summary of concluded results in absence of overload control mechanism

| output, without using overload control mechanism | 100cps | 300cps | 500cps | 700cps | 900cps | 1100cps | 1300cps | 1500cps | 1700cps |
|---|---|---|---|---|---|---|---|---|---|
| Call Setup delay (msec) | 6 | 10 | 12 | 100000 | 160040 | 180060 | 200000 | 200000 | 200000 |
| throughput(cps) | 100 | 300 | 500 | 673 | 90 | 70 | 50 | 30 | 20 |
| Retransmission rate(rps)-INVITE msg | 0 | 0 | 0 | 1500 | 5500 | 6500 | 7500 | 8500 | 9500 |
| Retransmission rate(rps)-BYE msg | 0 | 0 | 0 | 500 | 2000 | 2200 | 2400 | 2600 | 2800 |
| Avg memory(%) | 0 | 0 | 0 | 91 | 93 | 95 | 97 | 100 | 100 |
| CPU Utilization(%) | 28 | 36 | 60 | 92 | 100 | 100 | 100 | 100 | 100 |

Table 2: Summary of concluded results in presence of overload control mechanism

| output, with using overload control mechanism | 100cps | 300cps | 500cps | 700cps | 900cps | 1100cps | 1300cps | 1500cps | 1700cps |
|---|---|---|---|---|---|---|---|---|---|
| Call Setup delay (msec) | 6 | 10 | 12 | 15 | 30 | 40 | 50 | 150 | 18000 |
| throughput(cps) | 100 | 300 | 500 | 700 | 870 | 940 | 1000 | 1040 | 1020 |
| Retransmission rate(rps)-INVITE msg | 0 | 0 | 0 | 0 | 0 | 0 | 0 | 6000 | 8000 |
| Retransmission rate(rps)-BYE msg | 0 | 0 | 0 | 0 | 0 | 0 | 0 | 3000 | 3600 |
| Avg memory(%) | 0 | 0 | 0 | 0 | 0 | 0 | 0 | 90 | 100 |
| CPU Utilization(%) | 27 | 30 | 35 | 44 | 58 | 72 | 84 | 92 | 100 |

## 7. CONCLUSION

The studies accomplished in this paper show that SIP protocol is not efficient enough in facing with congestion, so that when call request rate increases, the delay of call establishment increases suddenly, proxy's throughput falls, and consequently resend rates and unsuccessful calls increase. In this paper window-based control method is developed, implemented, and tested on a real platform and also the efficiency of SIP proxy in case of overload is studied by using window-based distributed overload control method, which is developed on Asterisk open source proxy. Studying the charts of throughput, delay, and resend rate of INVITE and BYE messages in Asterisk proxy shows that the algorithm is able to maintain the throughput at up to twice of the downstream proxy's capacity. This is clearly observable in average memory and CPU utilization charts.